# The effect of Fe and Ru substitution for Ni on the thermopower in MgCNi$_3$

C. Sulkowski[1], T. Klimczuk[2,3], R. J. Cava[4], and K. Rogacki[1]

[1]Institute of Low Temperature and Structure Research, Polish Academy of Sciences,

P.O.Box 1410, 50-950 Wroclaw, Poland

[2]Faculty of Applied Physics and Mathematics, Gdansk University of Technology,

Narutowicza 11/12, 80-952 Gdansk, Poland

[3]Los Alamos National Laboratory, Los Alamos, NM 87545, USA

[4]Department of Chemistry, Princeton University, New Jersey 08544, USA

Abstract

The intermetallic perovskite MgCNi$_3$ is a superconductor with a T$_c$ = 7 K. Substitution of Fe and Ru for Ni decreases T$_c$ monotonically as the doping concentration is increased. Here we report thermopower measurements, S(T), on MgCNi$_3$, MgCNi$_{3-x}$Fe$_x$ and MgCNi$_{3-x}$Ru$_x$. For MgCNi$_3$, the thermopower is negative, - 12.5 µV/K, at 300 K. The absolute value of S decreases as x increases in MgCNi$_{3-x}$Fe$_x$ and MgCNi$_{3-x}$Ru$_x$. The sign of S changes from negative to positive at low temperatures for values of x > 0.01. These data show that the carriers in MgCNi$_3$ are electrons, and by increasing *x* and decreasing temperature, the participation of hole carriers clearly increases. The influence of the magnetic moments of the Fe atoms on the thermopower is not visible.

**Introduction**

MgCNi$_3$ is an unusual superconductor.[1] The high proportion of Ni atoms in the unit cell suggests the possibility of magnetic interactions, but so far this has not been observed. The simple intermetallic perovskite structure makes this compound attractive for study and there are many papers dedicated to theoretical considerations, i.e. band structure calculations. Analysis of the calculated electronic structure has shown a large narrow density of states peak located very close to the Fermi energy (E$_F$).[2,3,4,5,6,7,8] The presence of this peak was confirmed by photoemision and x-ray spectroscopy experiments.[6,9] Since the peak is located just below E$_F$, chemical substitution in MgCNi$_3$ is expected to significantly change its electronic properties. Numerous efforts have been made to hole dope MgCNi$_3$ in an attempt to shift the Fermi level thereby increasing the density of states at E$_F$. An increase in Tc or the appearance of ferromagnetism was expected. Previous studies have focused on the partial substitution of Co[10,11], Fe[10,12], Mn[13], and Ru[14] for Ni, the introduction of carbon deficiencies into the structure,[15,16] and on the partial substitution of B for C[17]. In all cases, Tc was found to decrease and ferromagnetism was not observed. Doping on the Mg site, which also causes a decrease of T$_C$, seems to be the most difficult (discussed in Ref. 18). Recently three new compounds in which Mg was completely replaced by Zn[18], Ga[19] and In[20] (GaCNi$_3$, ZnCNi$_3$ and In$_{0.95}$CNi$_3$) were reported.

Negative values for the Hall coefficient and thermopower indicate that the carriers in MgCNi$_3$ are electron-type.[21] The effect of Fe and Ru substitution for Ni on the superconducting transition temperature, T$_c$, in MgCNi$_3$ is quite different[12]. The superconductivity in MgCNi$_{3-x}$M$_x$ is suppressed more slowly in the Ru substituted compounds than in the Fe substituted compounds. This is most likely because the Fe atoms act as magnetic impurities that break the superconducting Cooper-pairs. Since Ru is a nonmagnetic metal, the observed changes in the $T_c$ of $MgCNi_{3-x}Ru_x$ are expected to be primarily due to a band structure (i.e., electron count) effect.[12] Therefore, studies of the transport properties of $MgCNi_3$ substituted with Fe and Ru are highly valuable. The elements Fe and Ru are from the same column in the periodic table and both substitutions are expected to decrease the electron concentration by the same amount. Measurement of the thermopower is a sensitive tool which can be used to monitor changes in the electronic properties of a material. In this communication, we report the results of our thermopower measurements, $S(T)$, on $MgCNi_3$, $MgCNi_{3-x}Fe_x$ and $MgCNi_{3-x}Ru_x$. We show that the Fe and Ru substitutions affect $S(T)$ similarly despite having a much different influence on $T_c$.

**Experimental**

Two series of 0.5 g samples with nominal compositions: $Mg_{1.2}C_{1.5}Ni_{3-x}Ru_x$ (x=0, 0.005, 0.01, 0.02, 0.03, 0.05 and 0.1) and $Mg_{1.2}C_{1.5}Ni_{3-x}Fe_x$ (x=0, 0.005, 0.01, 0.015, 0.02, 0.025, 0.05) were synthesized. The starting materials were Mg flakes (99% Aldrich Chemical), Ni sponge (99.9% Johnson Matthey and Alfa Aesar), glassy carbon spherical powder (Alfa Aesar), Ru powder (99.95% Alfa Aesar) and Fe powder (99.5% Alfa Aesar). Previous studies on $MgCNi_3$ indicated the need to employ excess magnesium and carbon in the synthesis in order to obtain optimal carbon content.[1,15] The excess Mg is vaporized during the course of the reaction (though it sometimes forms MgO in the final product)[15]. After thorough mixing, the starting materials were pressed into pellets, wrapped in zirconium foil, placed on an $Al_2O_3$ boat, and fired in a quartz tube furnace under a 95% Ar/5% $H_2$ atmosphere. The initial furnace treatment began with a half hour at 600 °C, followed by 1 h at 900 °C. After cooling, the samples were reground, pressed into pellets, and placed back in the furnace under identical conditions at 900 °C. The latter step was repeated two additional times. Following the heat treatment, the samples were analyzed with powder X-ray diffraction using CuKα radiation. The resulting material contains only one intermetallic phase, stoichiometric $MgCNi_{3-x}M_x$, plus a small proportion of elemental carbon.[15] Because no transition metal excess is needed for synthesis, and x in both $MgCNi_{3-x}Ru_x$ and $MgCNi_{3-x}Fe_x$ is far below the solubility limit, the nominal Ru and Fe contents correspond to the real doping level.

Superconductivity was characterized by zero field cooling AC magnetization ($H_{AC}$ = 3 Oe, f = 10 kHz) performed at 5 Oe DC field in the temperature range 1.9 - 8 K (PPMS—Quantum Design). Thermoelectric power measurements were performed in the temperature range 7-300 K by a steady-state-mode using a semiautomatic instrument fitted into the transport liquid helium dewar.[22] The sample was clamped between two spring-loaded Cu blocks with heaters attached. A pair of platinum thermometers (HY-CAL Engineering, EL-700-U, Pt-1000 Ω) was used to detect the temperature differences between the blocks. Special attention has been paid to limit any errors that might occur in the detection of small temperature differences. The blocks were insulated from the surroundings so that a thermal difference could be produced by the heaters. The quality of the thermal contact between the sample and the Cu blocks was tested by electrical resistance measurements and only values below 2 Ω were accepted. A calibration of the equipment was performed using a Pb (6N) sample.[23]

**Results and discussion**

The temperature dependence of the thermopower [S(T)] of MgCNi$_3$, MgCNi$_{3-x}$Ru$_x$ and MgCNi$_{3-x}$Fe$_x$ are presented in Figures 1 and 2, respectively. For MgCNi$_3$ the thermopower has a negative value in the temperature range 7-300 K and exhibits metallic character. The room temperature thermopower is S(300K) = - 12.8 µV/K and its magnitude is larger than that previously reported by Lin et al. (- 9.2 µV/K).[21] The absolute value of the thermopower, |S|, decreases as Fe and Ru are substituted for Ni. This indicates that changes in the density of states at the Fermi energy, $g(E_F)$, dominate over the influence of decreasing charge concentration, $n$, on the |S|. Assuming a constant value of $g(E_F)$, a decrease in $n$ should cause an increase in |S| according to the equation: $S(T) = \frac{2\pi^2 k_B^2}{3} \frac{g(E_F)}{n \cdot |e|} T$. However, |S| decreases with doping and this supports the suggestion that the $g(E)$ peak close to the Fermi Energy is smeared by elemental substitution, as a result, $g(E_F)$ decreases. The same conclusion was derived from the superconducting properties of the MgCNi$_{3-x}$M$_x$ (M=Fe, Ru). In this case, $T_c$ also decreases as the doping level increases.[12]

At low temperatures $S(T)$ changes sign from negative to positive and, for MgCNi$_{3-x}$Ru$_x$, this effect is visible in the concentration range: $0.01 \leq x \leq 0.05$. For high Ru concentrations, such as MgCNi$_{2.9}$Ru$_{0.1}$, $S(T)$ remains positive in the whole temperature range. Strong influence of the doping on S(T) is clearly visible in Figure 3 which shows the derivative of the thermopower with respect to temperature versus temperature (dS/dT vs. T). Above 50K, d$S$/d$T$ is negative and increases with temperature for MgCNi$_3$ and MgCNi$_{2.95}$M$_{0.05}$. Below 50K, the d$S$/d$T$ curve drops in the case of MgCNi$_3$ and rapidly increases in the doped samples. This opposite behavior indicates that the substitution of Fe or Ru causes large changes in the band structure of $MgCNi_3$. It also suggests a strong increase of hole participation in band conductivity.

Figure 4 illustrates the different effects of Fe and Ru doping of $MgCNi_3$ on the thermopower at 20K ($S_{20K}$) and the superconducting transition temperature ($T_C$). Superconductivity is suppressed more rapidly in the Fe substituted samples than in the Ru substituted samples (see inset). It is shown in Ref. [12] that magnetic susceptibility ($\chi$) in the normal state increases with Fe doping and decreases with Ru doping. This suggests that Fe acts as a magnetic impurity and breaks apart the Cooper-pairs. This effect was predicted by Abrikosov and Gorkov,[24] and was observed in many intermetallic superconductors.[25] The main panel of Figure 4 shows the thermopower at 20K for the Fe and Ru substituted samples. Interestingly, although $T_C$ decreases in a different way, the thermopower increases in a similar manner for both $MgCNi_{3-x}Ru_x$ and $MgCNi_{3-x}Fe_x$. The same effect is also observed in $Mg_{1-x}Mn_xB_2$ and $Mg_{1-x}Al_xB_2$ where Mg is partially substituted of by the magnetic atoms and the non-magnetic atoms.[26]

**Conclusions**

Our measurements of the temperature dependence of the thermopower, S(T), of $MgCNi_3$, $MgCNi_{3-x}Fe_x$ and $MgCNi_{3-x}Ru_x$ show that substitution Fe and Ru causes an increase of the participation of hole-type carriers. This effect is especially strong at low temperatures. The magnetic moments of the Fe atoms do not appear to have an effect on the S(T) of $MgCNi_{3-x}Fe_x$. This is in contrast to the strong dependence of the superconducting transition temperature on the Fe concentration. The thermopower changes greatly with Fe and Ru substitutions, especially at low temperatures. The Fermi energy in MgCNi$_3$ is located at the slope of $g(E)$. Therefore, it is expected that hole doping should increase both $T_C$ and $|S|$. Previous studies of MgCNi$_{3-x}$M$_x$, M = Co, Mn, Fe, Ru, and MgC$_x$Ni$_3$ and MgC$_{1-x}$B$_x$Ni$_3$ have shown the opposite effect, namely decreasing $T_c$. It is illustrated here that $|S|$ also decreases with the Fe and Ru doping. This supports the suggestion that the g(E) peak close to the Fermi Energy is smeared by elemental substitution and, as a result, $g(E_F)$ decreases.

**Acknowledgements**

Work at Los Alamos National Laboratory was performed under the auspices of the U.S. Department of Energy. Work at Princeton supported by grant DE-FG02-98-ER45706.

**Figure captions**

**Figure 1.** Temperature dependence of the thermopower, S(T), for all MgCNi$_{3-x}$Ru$_x$ samples, with x = 0, 0.005, 0.01, 0.02, 0.03, 0.05, and 0.1.

**Figure 2.** Temperature dependence of the thermopower, S(T), for all MgCNi$_{3-x}$Fe$_x$ samples with x = 0, 0.005, 0.01, 0.025, and 0.05.

**Figure 3.** Derivative d$S$/d$T$ vs. temperature for MgCNi$_3$ and MgCNi$_{2.95}$Fe$_{0.05}$, and MgCNi$_{2.95}$Ru$_{0.05}$.

**Figure 4.** (color on-line) Thermopower at 20K, $S_{20K}$, and the superconducting transition temperature, $T_c$, (inset) in MgCNi$_{3-x}$M$_x$ (M = Ru, Fe) as a function of doping, $x$.

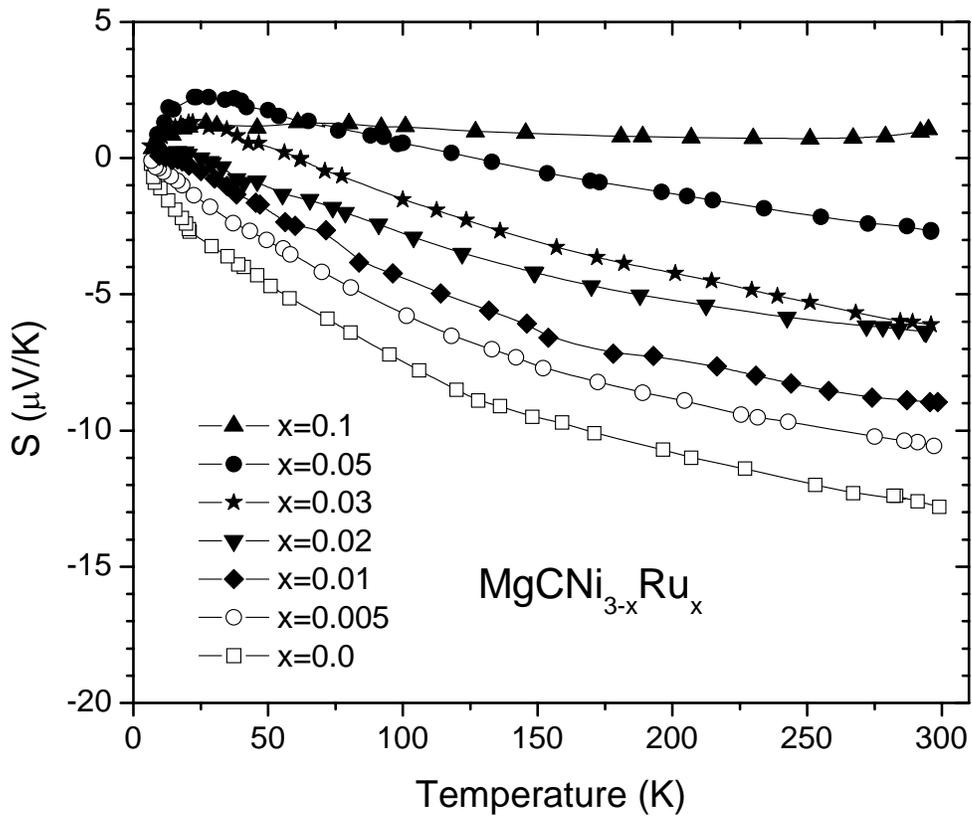

Figure 1.

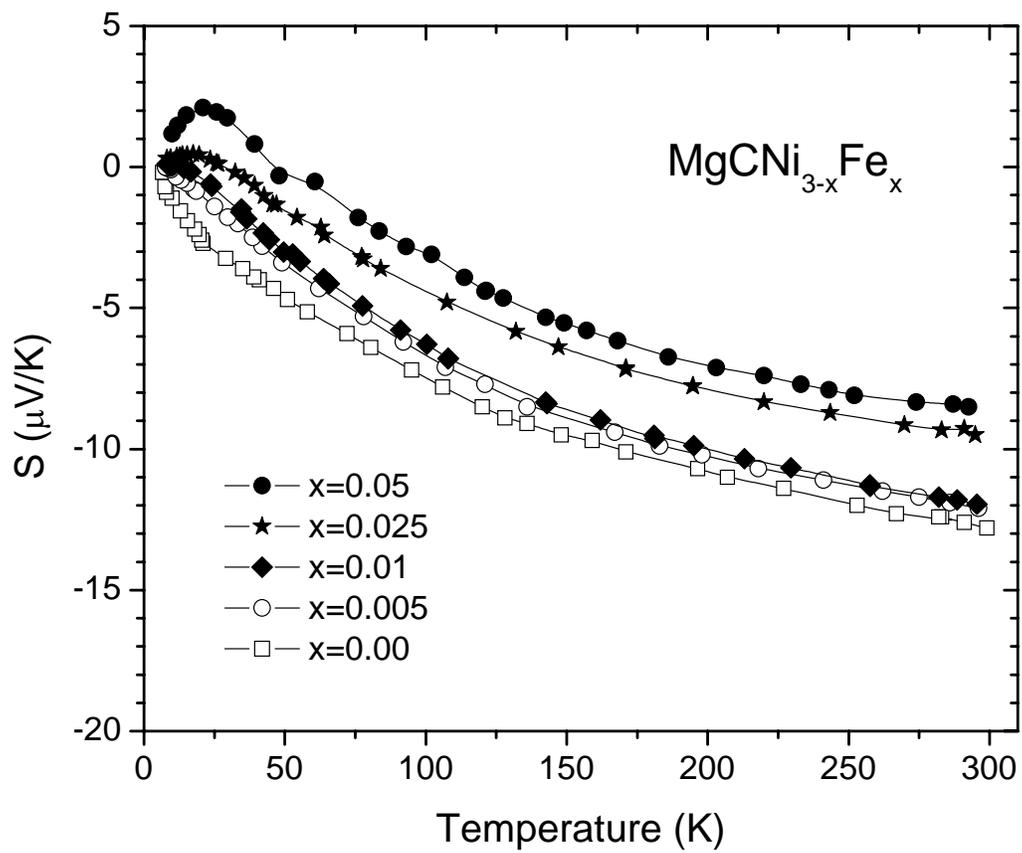

Figure 2.

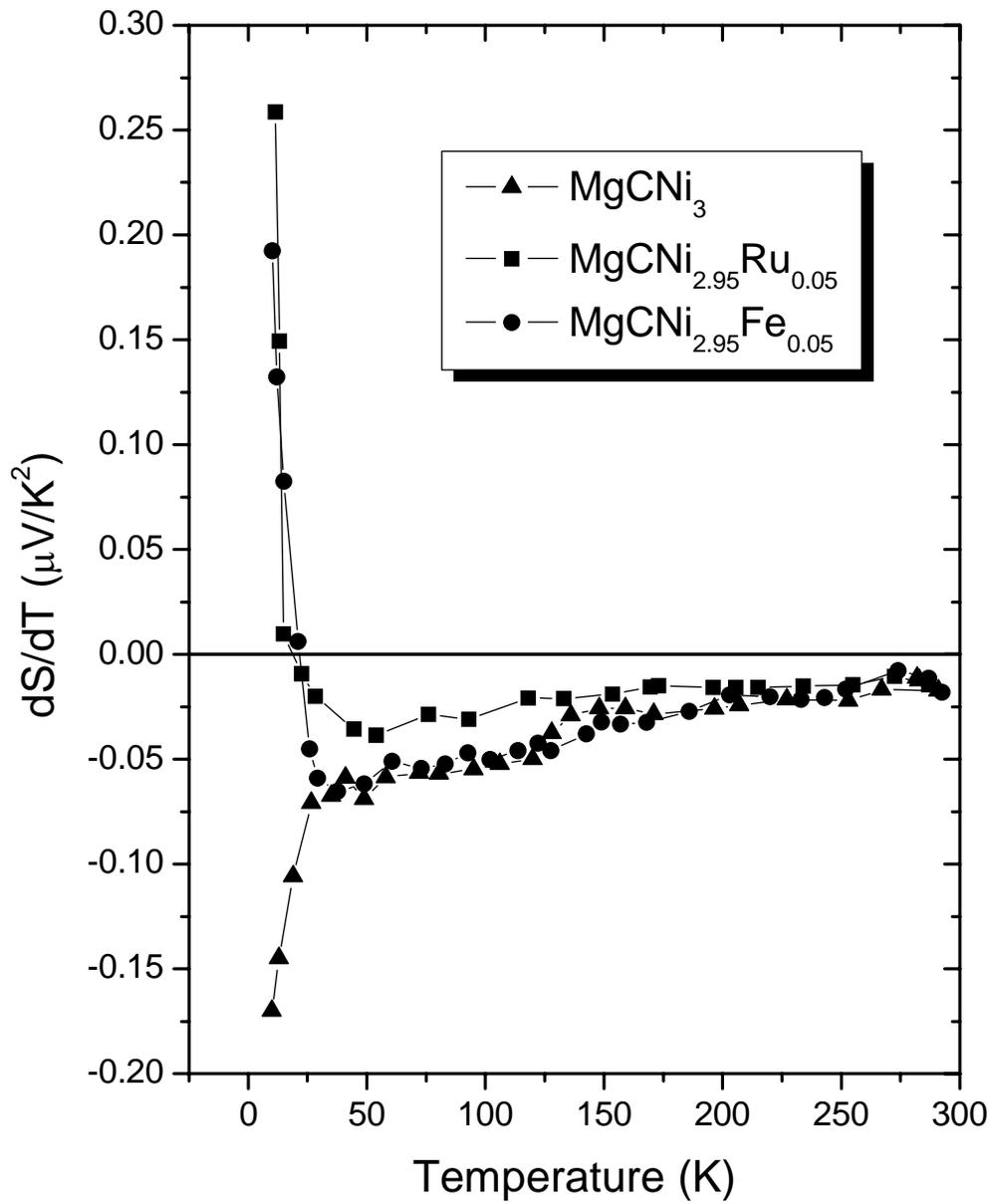

Figure 3.

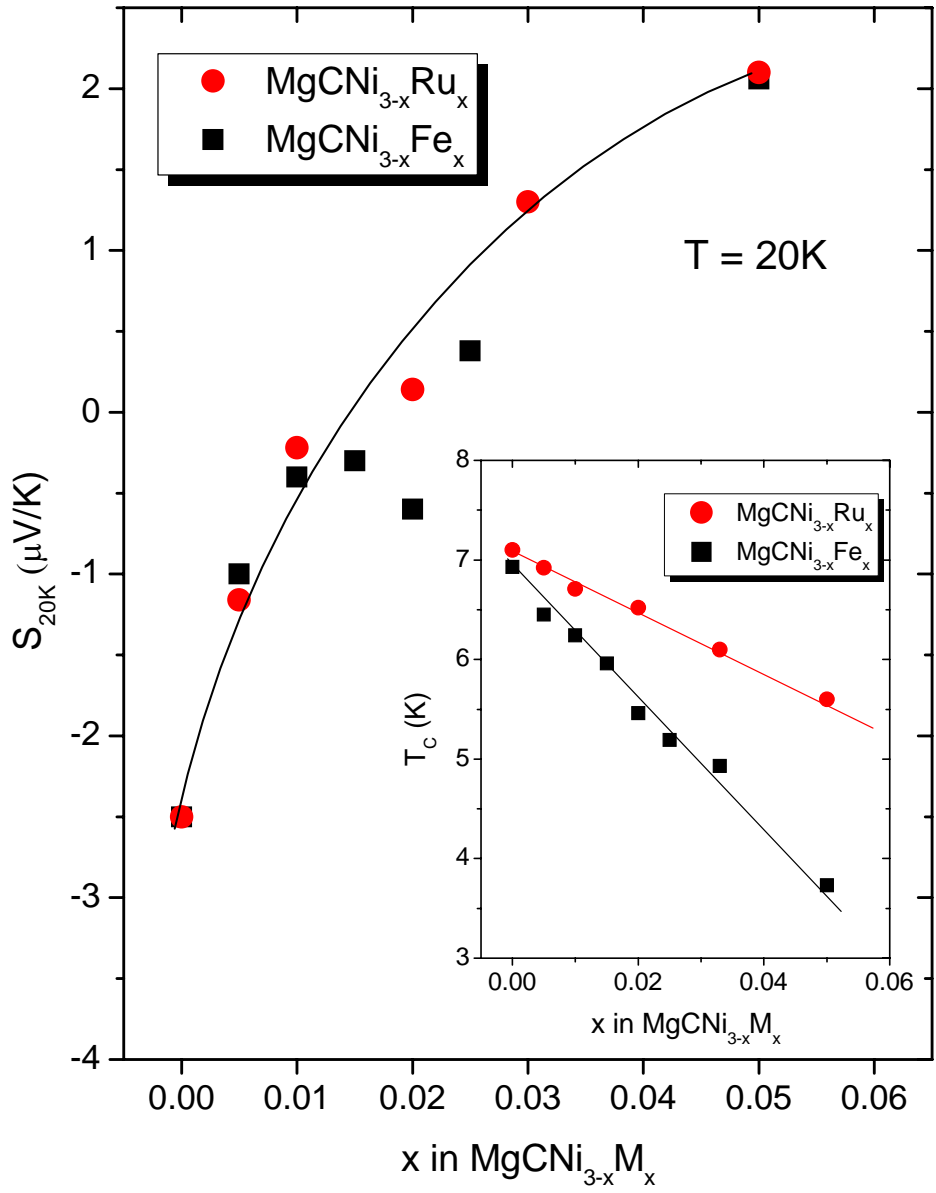

Figure 4.